\documentclass[12pt,a4paper]{article}
\usepackage{graphicx}
\usepackage{amssymb}

\begin{document}

\title{\bf Dependence of vertical cutoff rigidities and magnetospheric
transmission on empiric parameters}

\author{Vladimir V. Petrukhin\footnote{vovka@dec1.sinp.msu.ru}\\
\small Skobeltsyn Institute for Nuclear Physics, Moscow State University\\
\small 119991, Moscow, Leninskie gory, 1(2)
}

\date{}

\maketitle

\begin{abstract}
Using dynamic paraboloid model of Earth's magnetosphere,
a large set of particles' trajectory computations was
performed. Based on its result, the numerical algorithm
for calculating effective cutoff rigidity dependence on
empiric parameters has been developed for further use in
magnetospheric transmission calculations.
\end{abstract}

{\it keywords: cutoff rigidity, dynamic paraboloid model, magnetospheric
transmission, radiation condition on LEO}


\section{Introduction}
Radiation condition onboard LEO spacecrafts is determined,
particularly, by the charged particles penetrating from outside
of the magnetosphere. The measure of such a process in any
given near-Earth space location is a local geomagnetic cutoff
rigidity value. In this work we assume, that "cutoff rigidity"\ 
phrase denotes an effective vertical cutoff rigidity
($R_{eff}$), which value is gained from calculated discrete
penumbra structure by ordinary technique using white
spectra (for example, ~\cite{eff}).

It is well known, that magnitude of $R_{eff}$ and penetration
boundaries' position depend on local time and different
magnetospheric condition-related parameters
~\cite{lt1,lt2,lt3,cond1,cond2,sm1,sm2}.
Cutoff rigidity value's variations can be measured in experiments
or obtained by some type of computation using Earth's magnetosphere
model. However, usual method for $R_{eff}$ calculation is
based on resource consuming trajectory computation technique
~\cite{traj1,traj2}.

In this work we offer the method to calculate $R_{eff}$ with
accounting for local time and main empirical parameters,
that characterizes magnetospheric condition, in any point
of near-Earth's space. For the applications where $R_{eff}$
calculation is often needed (e.g. transmissions for LEOs)
our method provides a huge speedup in comparison with direct
trajectory computations. The only things needed for the method
are the IGRF rigidity in exploring point and rigidity attenuation
~\cite{nym1} quotient's dependence on empirical parameters.
The first one can be obtained using interpolation in
$R_{eff}^{IGRF}$ tabulated data (see, for example,
~\cite{sm3,nym2}), which is regularly updated for coming epochs.
And the second
can be obtained only once by using particle trajectory
computations in corresponding numerical magnetospheric model.
In our case the dynamic paraboloid model ~\cite{p1,p2,p3}
of Earth's magnetosphere was chosen for solution of the problem.

\section{Cutoff rigidity calculation scheme}
We have adopted in our work the cutoff rigidity attenuation
formalism ~\cite{nym1,nym2}, which has already successfully
applied ~\cite{creme} for transmission calculations using
Tsyganenko-89 ~\cite{t89} model. The rigidity attenuation
quotient $\Delta$ in our case is calculated as
 $$\Delta^{model}=R_{eff}^{model}(quiet)/R_{eff}^{model}(disturbed)$$
where $R_{eff}^{model}(quiet)$ value corresponds the set of
model parameters for non-disturbed magnetosphere condition,
and $R_{eff}^{model}(disturbed)$ is correspond to some set
of varied parameters regarding to quiet ones.

As it was wordlessly postulated in ~\cite{nym1}, the $\Delta$
dependence on point's geographical position might not be taken
into account because it is relatively small. Generally, it
seems to be true at least for $R \gtrsim 0.6$ GV with good enough
accuracy (error's order is about percents even for lowest $R$
in this range). Hence, the approach of calculating a world-wide
$R_{eff}$ grid can be essentially simplified.

The algorithm we propose for obtaining $R_{eff}^{model}(LT, parameters)$
in given point contains 3 stages:
\begin{itemize}
\item Calculating $R_{eff}^{IGRF}$ for the given point; for
practical needs, the interpolation in pre-calculated grid for some
altitude $H_0$ is often used, it is especially easy when large amount
of results is needed quickly. The
formula
$$ R_{eff}^{IGRF}(H) = R_{eff}^{IGRF}(H_0) \cdot
 \left( \frac{r_E + H_0}{r_E + H} \right) ^ {2} $$
can be applied to transform $R_{eff}^{IGRF}$ value to new altitude $H$.
\item Calculating a model-dependent cutoff rigidity attenuation
 quotient \\
 $\Delta^{model}(R_{eff}^{IGRF}, LT, parameters) =
  R_{eff}^{model}(quiet)/R_{eff}^{model}(R_{eff}^{IGRF}, LT, parameters)$
 by interpolation in pre-calculated $R_{eff}^{model}$ database
(or by extrapolation if $R_{eff}^{IGRF}$ value is out of applied
 basic $R_{eff}^{IGRF}$ array, see Appendix)
\item Computing the resulting value $R_{eff}^{model} =
 R_{eff}^{IGRF}/\Delta^{model}(R_{eff}^{IGRF}, LT, parameters)$
\end{itemize}
Table of $R_{eff}^{IGRF}$ values for some given altitude $H_0$
is needed to be renewed every 5 years with new IGRF epoch coming.
Values $R_{eff}^{model}(quiet)$ were obtained by trajectory
computations using dynamic paraboloid model with parameter set
$[7, 380, 0, 0, 5, -3, 1]$ (sequence of variables is according
to table \ref{t1} below).
To compute $R_{eff}^{model}$ database which is intended to provide
dependence of $\Delta^{model}$ on parameters and local time $LT$,
we used the varying of all model parameters in wide enough ranges,
comprising extreme, quiet and "anti-extreme"\ parameter sets. The
multidimensional grid, applied for this calculation, is presented
in table \ref{t1}.
\begin{table}[h]
\caption{The applied parameter grid.}
\label{t1}
\centering
\begin{tabular}{c|cccc}
\hline
$\rho$ & 0.5 &2 &8 &20        \\
$v$ & 400 &700 &1200 &2000    \\
$Dst$ & -460 &-150 &-50	&0    \\
$AL$ & -2000 &-800 &-200 &0   \\
$B_x^{IMF}$ & -40 &-10 &5 &20 \\
$B_y^{IMF}$ & -25 &-5 &20 &50 \\
$B_z^{IMF}$ & -50 &-20 &5 &30 \\
\hline
\end{tabular}
\end{table}
Values at the edges of this grid were obtained (approximately)
from ~\cite{kal}. All calculations were performed for local time
values $3, 9, 15$ and $21$ hours. Technical details of calculations
are summarized in Appendix.
\begin{figure*}
\centering
\includegraphics[width=\textwidth]{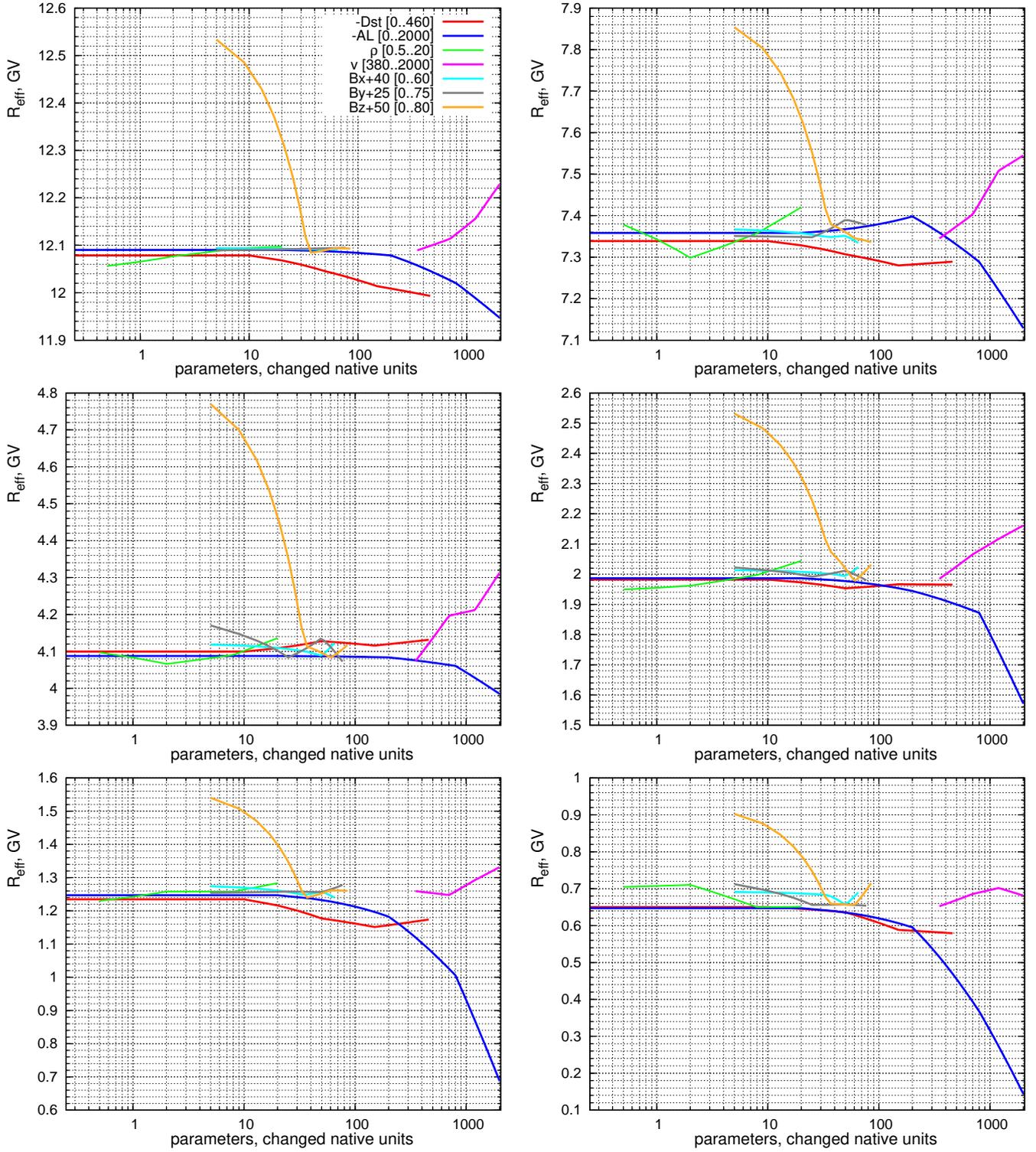}
\caption{Partial dependencies of $R_{eff}$ on model parameters
 for six explored points.}
\label{f1}
\end{figure*}
Let us now demonstrate some results of cutoff rigidity dependence
on local time and model parameters, that have been computed using
presented technique.

\begin{figure*}[h]
\centering
\includegraphics[width=0.9\textwidth]{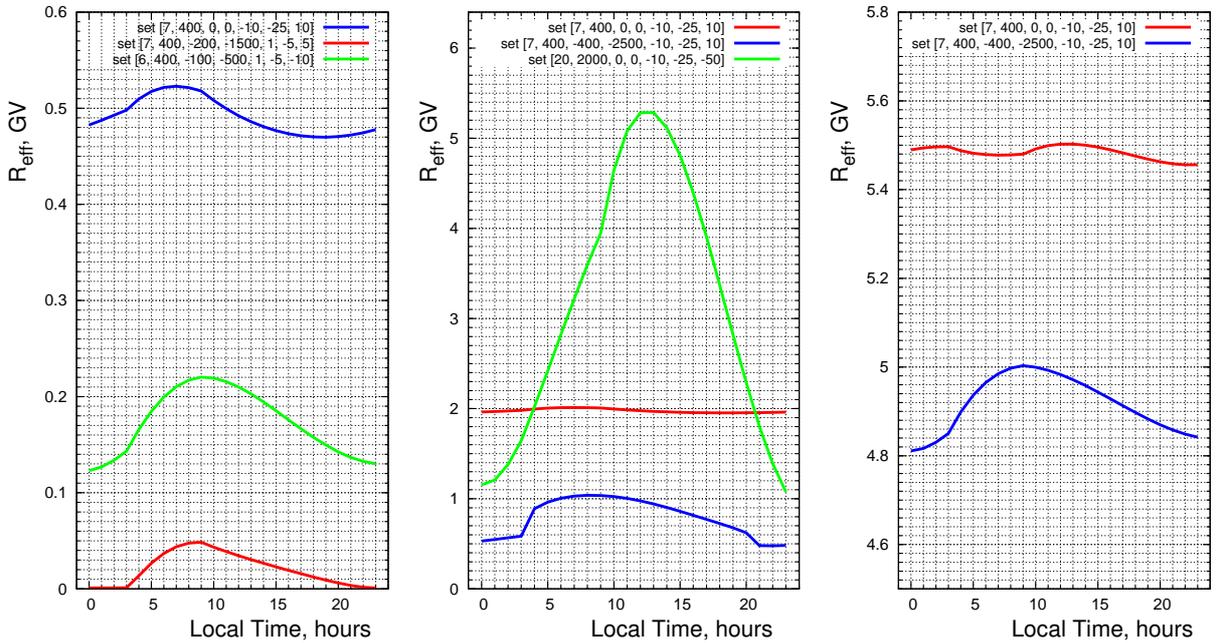}
\caption{Dependencies of $R_{eff}$ on local time and model
parameter set for three points with $R_{eff}^{model}(quiet)$
values 0.49 GV (left), 1.98 GV (middle) and 5.48 GV (right).}
\label{f2}
\end{figure*}

Fig.\ref{f1} gives an opportunity to weigh deposit of every model
parameter in resulting $R_{eff}$ value for all points of noted
above set. Such case of partial dependencies is obtained by
fixating all parameters as quiet and excepting the given one.
Note, that the role of $Dst$ as $R_{eff}$ depression factor wanes
with decreasing of $R_{eff}$, but the $AL$ one rises rapidly.
The weight of $B_x^{IMF}$ and $B_y^{IMF}$ seems to be not
important for $R_{eff}$ values, at any hand, in presented
case of quiet basic values, because they result in $R_{eff}$
changes that are comparable with the value of applied rigidity
step size (see Appendix). Contrary, decrease of $B_z^{IMF}$ effects
in huge $R_{eff}$ grow (see also fig.\ref{f2}). The effect of
$\rho$ and $v$ here is clear but of moderate magnitude.

Fig.\ref{f2} shows family of $R_{eff}$ daily variations for three
points having normal cutoff values of different order. For the first
picture there are some hours where $R_{eff}$ is negligibly small,
so such a point can be accessible for particles of any rigidity.
The middle picture demonstrates the effect of
$B_z^{IMF}$\ "anti-extreme"\ setting, leading not only to very
high $R_{eff}$ mean values (for this case), but also to the
translocation of curve's maximum for time axis's positive
direction. The upper curve on the most right part of Fig.\ref{f2}
exhibits the insufficiency of applied step size for penumbra
calculation (see Appendix), leading to the underestimation of
$R_{eff}$ value for $LT=9$, which causes an ambiguous
interpolation results. However, this effect is rare and too
weak to affect resulting transmission, for which calculating
presented method was created.

A "gnarlyness"\ of some presented curves is due to effect of
scarce grid, currently imperfect interpolation procedure and
scale. It is does not sufficiently affect the results of applications
where much of $R_{eff}$ values to be calculated, because
interpolation uses many $R_{eff}^{model}$ values and their errors
(which are of order of $R$ step size) are to be evened. Extrapolating
parameter values to out of grid (but not very far) is also possible
and seems to be correct enough to get reliable result.

\section{Transmission functions for LEO missions}
\begin{figure*}[h]
\centering
\includegraphics[width=0.9\textwidth]{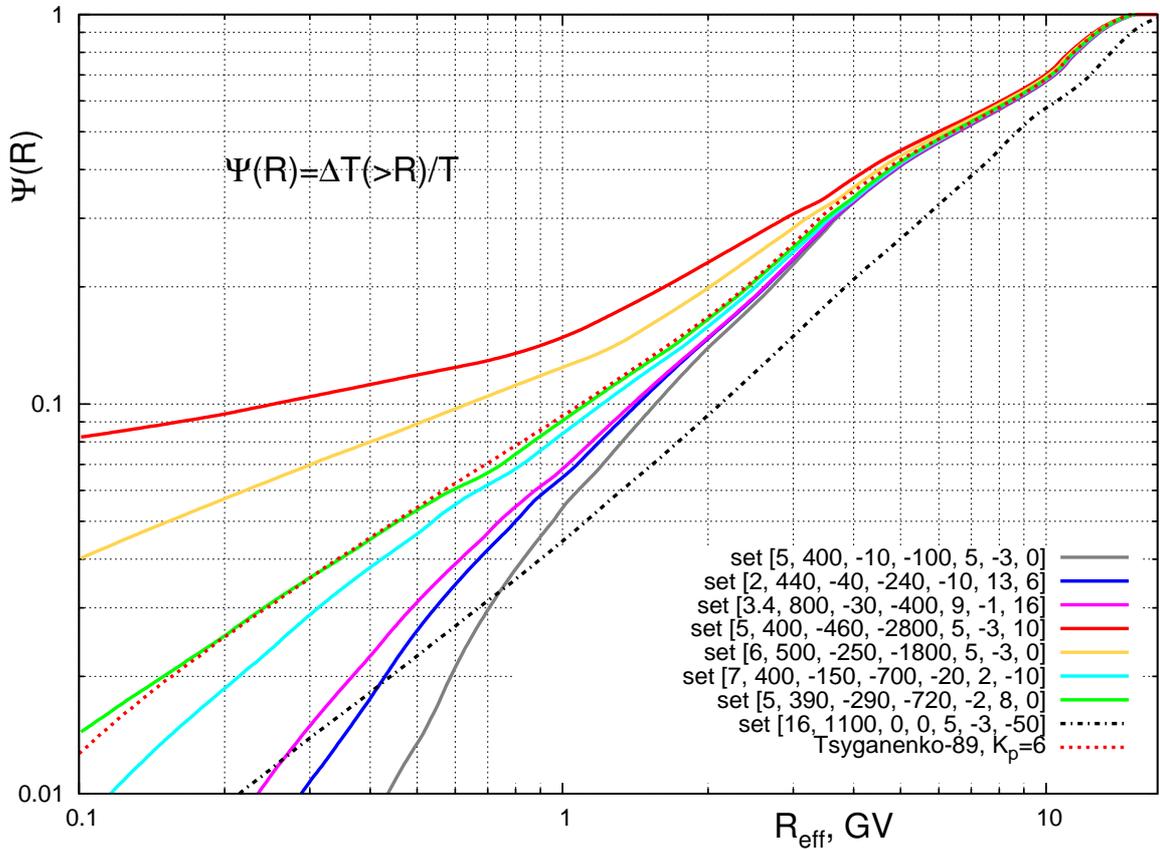}
\caption{Magnetospheric transmission $\Psi(R)$ for different conditions,
obtained according to paraboloid model with using presented technique.
Tsyganenko-89 transmission for $Kp=6$ is also figured here for
comparison. Note also the standing out of all "anti-extreme"\ case.}
\label{f3}
\end{figure*}

Magnetospheric transmission functions are directly connected
with such tasks as LEO radiation conditions evaluating and
the interpretation of orbital experiment results~\cite{trans}.
Because of their importance, it is necessary to be able to
account for more empiric factors that affecting transmission.
Here we'll give some examples of calculated magnetospheric
transmission functions for several (quiet, disturbed and very
extreme) model parameters sets for ISS-like orbit, obtained using
presented technique. The essential note is that, unlike Tsyganenko
models, the paraboloid one is primordially intended to describe
even very extreme conditions, as it is not internally limited by
experimental data arrays. Fig.\ref{f3} presents some modeled
transmissions for $H=450$ km $i=51.7^o$ circular orbit under
different conditions for long mission duration, where all localities
are averaged and transmission curve became smoothed. In comparison
with additionally figured transmission, obtained by Tsyganenko-89
(T89) model with $Kp=6$, it is obviously seen that the paraboloid
model gives unapproachable for T89 results in modeling most extreme
conditions.

\section{Conclusion}
The method we have developing is intended for applications
using intensive geomagnetic cutoff calculations, first of all
for express magnetospheric transmission calculation for given LEO
missions, even under changing magnetospheric conditions during flight.
Test $R_{eff}$ and transmission calculation, based on the
presented technique, is available online by URL
http://dec1.sinp.msu.ru/\symbol{126}vovka/riho
(a simplified version, where transmission calculation is allowed
for only static conditions and circular orbits).

Current version of the method uses the IGRF rigidity values table
as basic (in step 1 of scheme), although it seems that the best
choice would be to apply some other world basic grid, for example,
one directly calculated $R_{eff}^{model}(quiet)$ table with using
paraboloid model.
Nevertheless, it does not understate the attenuation quotient
methodology itself, because its drawback in our case is only
correspond to relatively small systematic inaccuracies between
conditional "quiet"\ $R_{eff}^{model}$ values and IGRF ones.

\section*{Acknowledgement}
Author thanks Dr.Nymmik for fruitful discussions on the topic
of paper, and Dr.Kalegaev for the provided data and cluster
computing facilities.

\section*{Appendix. Technical details}
Here we summarize technical moments that are related to the
performed $R_{eff}^{model}$ trajectory calculations with
using paraboloid model.

Vertically directed protons were test particles for reverse
trajectory calculations of penumbra. Fourth order integration
scheme was used for it, with accounting for time changing
(hence, magnetic field) during particle's modeled flight.
The geomagnetic field there was superposition of IGRF epoch
2005 field with dynamic paraboloid model (version 2004
~\cite{parinet}), with uniform field vector
$\overrightarrow{B^{IMF}}$
outside of the magnetopause, which position is natively given by
the code realizing the paraboloid model. Maximal flying distance
was equal to $15 r_E$, the particles walked it over during
motion were considered as allowed for penetration. The Earth
was represented by WGS-82 ellipsoid with atmosphere layer at
$20$ km above its surface, the particles which fell below it was
considered as reentrant. Rigidity step size for penumbra calculation
was equal to $0.01$ GV. All calculations were performed for
altitude $450$ km above mean Earth's radii $r_E=6371.2$ km for
six selected points in northern geographic hemisphere with basic
IGRF rigidities $12.086, 7.381, 4.101, 2.018, 1.203, 0.666$ GV.

All presented transmissions were obtained for mission duration
3000 revolutions with orbital trajectory step size 0.9 degree.

\end{document}